\documentclass[twocolumn,showpacs,preprintnumbers,amsmath,amssymb]{revtex4}

\usepackage{graphicx}
\usepackage{dcolumn}
\usepackage{bm}
\usepackage{amsmath,dsfont}

\newcommand{\half}{{\scriptstyle{\frac{1}{2}}}}
\newcommand{\const}{\mathop{\rm const}\nolimits}

\def\IR{{\mathds{R}}} 

\def\p{{\partial}}
\def\vnabla{{\vec{\nabla}}}

\def\vp{{\vec{p}}}
\def\vx{{\vec{r}}}
\def\hx{{\hat{r}}} 
\def\vn{{\hat{n}}}
\def\vm{{\hat{m}}}
\def\vJ{{\vec{J}}}
\def\vL{{\vec{L}}}
\def\vK{{\vec{K}}}
\def\vI{{\vec{I}}}
\def\vA{{\vec{A}}}
\def\vB{{\vec{B}}}
\def\vD{{\vec{D}}}
\def\vC{{\vec{C}}}
\def\vX{{\vec{X}}}

\def\cD{{\cal{D}}}
\def\vW{{\vec{\Psi}}}

\def\vPsi{{\vec{\Psi}}}
\def\vpi{{\vec{\pi}}}
\def\vN{{\vec{N}}}
\def\su{{\rm su}}
\def\SU{{\rm SU}}

\def\SO{{\rm SO}}
\def\smallover#1/#2{\hbox{$\textstyle\frac{#1}{#2}$}} 

\begin{document}

\preprint{arxiv:0902.0273}

\title{Conserved quantities in non-abelian monopole fields}

\author{\large P.~A.~Horv\'athy and J.-P. Ngome}
\affiliation{
Laboratoire de Math\'ematiques et de Physique Th\'eorique, 
Universit\'e de Tours.
Parc de Grandmont. 
F-37200 Tours (France)}
\email{horvathy-at-lmpt.univ-tours.fr, 
ngome-at-lmpt.univ-tours.fr .}

\date{\today}

\begin{abstract}
Van Holten's covariant Hamiltonian framework is used to find conserved quantities for an isospin-carrying particle in a non-Abelian monopole-like field. For a Wu-Yang monopole we find the most general scalar
potential such that the combined system admits a conserved 
Runge-Lenz vector. It generalizes the fine-tuned inverse-square 
plus  Coulomb potential, found before by McIntosh and Cisneros, and by Zwanziger, for a charged particle in the field of a Dirac monopole.
Following Feh\'er, the result is interpreted as describing motion in the asymptotic field of a self-dual Prasad-Sommerfield monopole. 
In the effective non-Abelian field for 
 nuclear motion in a diatomic molecule due to Moody, Shapere and Wilczek, 
 a conserved angular momentum is constructed, despite the non-conservation of the electric charge. No Runge-Lenz vector has been found. 
\end{abstract}


\maketitle

\section{Introduction}
In a recent paper van Holten \cite{vHolten}
outlined an algorithm for deriving conserved
quantities for a particle in a given external field, based on using
Killing tensors. An example illustrating his method
is provided by an isospin-carrying
particle \cite{Wong,Bal,Duval:1978,JaMa,DHInt} in the field
of a Wu-Yang monopole \cite{WuYang,WYmotion}. He concluded, in particular, that no Runge-Lenz type
vector could exist for such a system.
 This result reminds one to an earlier result  by Feh\'er
\cite{Feher:1987}, who proved that a charged particle in a Dirac monopole field can 
not have a globally defined Runge-Lenz vector.
In the latter case there is a way out, though  
\cite{MICZ,Schonfeld,Feher,Plyush,Feher:1987}~: adding a fine-tuned inverse square potential removes 
the obstruction, providing us with a conserved Runge-Lenz vector.

Below we  find, using van Holten's recipe, 
the most general additional potential
which allows for a conserved Runge-Lenz vector in non-Abelian monopole-like fields. 

In the Wu-Yang case, the resulting combined system has a remarkable
physical interpretation: it describes the motion in the long-distance limit 
of a self-dual monopole \cite{Feher}. 

Similar results hold in the effective
field of a diatomic molecule, considered before by Moody, Shapere and Wilczek \cite{MSW,Jdiat,Zyg}. Despite the non-conservation of the electric charge, we can construct a conserved angular momentum. No Runge-Lenz vector is found in general, though.

\section{Conserved quantities}

We start with the 
equations of motion for an isospin-carrying particle in 
a static non-Abelian gauge field \footnote{
Both the  gauge field strength
$F_{\mu\nu}=F^a_{\mu\nu}T_a,\ F_{jk}^a=\partial_jA_k^a-\partial_k\,A^a_j
-\epsilon_{abc}\,A_j^bA_k^c$,  and the classical isospin, $I=I^aT_a$,
take their values in the Lie algebra. For simplicity we restrict our attention
to 
$\su(2)\approx\IR^3$ so that $T_a=\half\sigma_a$.
$\mu,\nu\dots$ are space-time indices, $i,j,$ etc are space indices, and $a=1,2,3$
internal index referring to the non-Abelian Lie algebra generators $T_a$.
The ``dot'' denotes differentiation
w.r.t. non-relativistic time, $t$, along the 
trajectory.}
\begin{equation}\left\{
\begin{array}{lll}
\dot{\pi}_i&=&I^aF^a_{ij}\dot{x}^j-D_iV,
\\[8pt]
\dot{I}^a&=&-\epsilon_{abc}I^b\big(A_j^c\dot{x}^j-\displaystyle\frac{\p V}{\p I^c}\big),
\end{array}\right.
\label{Wongeq}
\end{equation}
where $\pi_i=\dot{x}_i$,
 and a scalar potential, $V=V(\vx,\vI)$ has also  been included for later convenience. 
Defining the
covariant Poisson bracket and Hamiltonian as \cite{vHolten},
\begin{eqnarray}
\big\{f,g\big\}&=&D_jf\frac{\p g}{\p \pi_j}-\frac{\p f}{\p \pi_j}D_jg 
\label{PBracket}
\\[4pt]
&&+\;I^aF_{jk}^a\frac{\p f}{\p \pi_j}\frac{\p g}{\p \pi_k}
-\epsilon_{abc}\frac{\p f}{\p I^a}\frac{\p g}{\p I^b}I^c,
\nonumber
\\[6pt]
H&=&\frac{1}{2}\vpi^2+V(\vx,\vI),
\label{cHam}
\end{eqnarray}
where $D_j$ is the covariant derivative,
\begin{equation}
D_jf=\p_jf-\epsilon_{abc}I^aA_j^b\frac{\p f}{\p I^c}.
\label{covder}
\end{equation}
Let us record the commutation relation of the
covariant derivatives,
\begin{equation}
[D_i,D_j]=-\epsilon_{abc}I^aF_{ij}^b\frac{\partial}{\partial I^c}\ . 
\label{covdercomm}
\end{equation}

The equations (\ref{Wongeq}) can be obtained in a Hamiltonian framework,
$\dot{x}_i=\big\{x_i,H\big\}, \ \dot{\pi}_i=\big\{\pi_i,H\big\}, \ \dot{I}^a=\big\{I^a,H\big\}$.

Following van Holten \cite{vHolten},
constants of the motion can conveniently be sought for in the form of an expansion into powers of the covariant momentum,
\begin{equation}
Q=C(\vx,\vI)+C_i(\vx,\vI)\pi_i+\frac{1}{2!}C_{ij}(\vx,\vI)\pi_i\pi_j+\dots
\label{consquant}
\end{equation}
Requiring $Q$ to Poisson-commute with the Hamiltonian yields a series
of constraints,  
\begin{equation}
\begin{array}{llll}
C_iD_iV&=&0,&\hbox{order 0}
\\[8pt]
D_iC&=&I^aF^a_{ij}C_j+C_{ij}D_jV,
&\hbox{order 1}
\\[8pt]
D_iC_j+D_jC_i&=&I^a(F^a_{ik}C_{kj}+F^a_{jk}C_{ki})&
\\[6pt]
&&\;+\;C_{ijk}D_kV,
&\hbox{order 2}
\\
\vdots&&\vdots&\vdots
\end{array}
\label{constraints}
\end{equation}
The expansion can be  truncated at a finite order, provided the covariant Killing equation is satisfied,
\begin{equation}
D_{(i_1}C_{i_2\dots i_n)}=0,
\label{Killingcond}
\end{equation}
when we can set $C_{i_1\dots i_{n+1}\dots}=0$.
For $n=1$, (\ref{Killingcond}) is a Killing vector. For example, we have, for any unit vector $\vn$,  the generator of a rotation around $\vn$,
\begin{eqnarray}
\vC=\vn\times\vx,
\label{rotKilling}
\end{eqnarray}
Then  
van Holten's recipe yields conserved angular momentum.
Similarly, for any unit vector $\vn$,
\begin{equation}
C_{ij}=2\delta_{ij}\, \vn\cdot\vx-(n_ix_j+n_jx_i)
\label{RLKilling}
\end{equation}
is a Killing tensor of order $2$, associated with the Runge-Lenz vector of planetary motion.

\section{Wu-Yang monopole}\label{WYSec}

The prototype of non-Abelian monopoles is  the one due to
Wu and Yang \cite{WuYang},  given by 
the non-Abelian gauge potential with a ``hedgehog'' magnetic field,
\begin{eqnarray}\displaystyle{
A_i^{a\;WY}=\epsilon_{iak}\frac{x_k}{r^2}}\ ,
\quad
F_{ij}^a =\epsilon_{ijk}\,\frac{x_k\;x_a}{r^4}\ .
\label{WYmonopA}
\end{eqnarray}

Let us now consider an isospin-carrying particle moving in a Wu-Yang monopole field augmented by a scalar potential, and inquire about
conserved quantities.

To start, we search for a conserved quantity of order zero, i.e.,
$C_i=C_{ij}=\dots=0$. Then (\ref{constraints}) is satisfied for an arbitrary potential if 
$D_iC=0$. In particular, 
\begin{eqnarray}
Q=\vI\cdot{\hx}
\qquad(\hx=\frac{\vx}{r}),
\label{echarge}
\end{eqnarray}
is covariantly constant, $\vD Q=0$, and $Q$ is, therefore, a constant of the motion. An easy calculation shows, furthermore, that 
(\ref{echarge}) is the only such quantity.

Next, we study conserved quantities which are linear in $\pi_i$,
$C_{ij}=\dots=0$. 
When the potential is invariant w.r.t. joint rotation of $\vx$ and $\vI$ , inserting the Killing vector 
(\ref{rotKilling}) into
equations (\ref{constraints}) yields \footnote{We stress that here and
in the subsequent formulas, $Q$ is \emph{not} a constant, merely
the specific combination (\ref{echarge}), which happens to be conserved along the trajectories of our particle.},
\begin{equation}
C=-\vn\cdot(Q\,{\hx}).
\label{WYC}
\end{equation}
and we end up with the angular momentum \cite{vHolten},
\begin{eqnarray}
\vJ=\vx\times\vec{\pi}-Q\,{\hx}\;.
\label{WYangmom}
\end{eqnarray}

Let us now turn to quadratic quantities. We observe that
the Killing tensor  has the property,
 
\begin{equation}
C_{ij}x_j=(\vn\cdot\vx)x_i-r^2n_i=-r^3\p_i\left(\vn\cdot{\hx}\right),
\label{radCij}
\end{equation}
Inserting (\ref{RLKilling}) into (\ref{constraints}), from the 2nd-order equation we find, therefore,
\begin{eqnarray}
\vC=\vec{n}\times(Q\,{\hx})\;. 
\label{WYvC}
\end{eqnarray}
Restricting ourselves to potentials with fall off at infinity,
$
V=\sum_{m=0}^\infty\alpha_m\,r^{-m},
$
the zeroth order equation allows us to infer that the real coefficients $\alpha_m$ are covariantly constant. 
The first-order equation requires in turn
\begin{eqnarray}
D_iC = \Big[\frac{Q^2}{r}-\sum_{m=0}^N m\;\alpha_m\,r^{-m+1}\Big] 
\,r^3\p_i\left(\vn\cdot{\hx}\right).
\label{WYCeq}
\end{eqnarray}
Our constraint can be solved therefore if $N=2$ and the leading coefficient in the expansion of $V$ is chosen to cancel the obstruction term
$Q^2/r$. Hence,
\begin{equation}
V=\frac{Q^2}{2r^2}+\frac{\alpha}{r}+\beta
\quad\hbox{and}\quad
C=\alpha\;\vec{n}\cdot{\hx},
\label{goodpot}
\end{equation}
where $\alpha$ and $\beta$ are arbitrary constants. Note that
the coefficient of the inverse-square term is fixed by the requirement of canceling the $Q^2/r^4$ term in the first bracket. 
Collecting our results yields,
\begin{eqnarray}
\vK=
\vec{\pi}\times{\vJ}+\alpha\,{\hx}\,,
\label{WYRL}
\end{eqnarray}
which is indeed a conserved  Runge-Lenz vector for an isospin-carrying particle in the  
Wu-Yang monopole field, combined with the potential (\ref{goodpot}) \cite{HWY}.

The importance of the conserved quantities $\vJ$ and $\vK$ is understood by noting that they determine the trajectory~: multiplying (\ref{WYangmom}) by the position, $\vx$, yields
\begin{equation}
\vJ\cdot{\hx}=-Q,
\label{cone}
\end{equation}
so that the particle moves, as always in the presence of a monopole,  on the surface of a cone of 
half opening angle $\theta={\rm arccos}(|Q|/J)$ ($J=|\vJ|$).

On the other hand, the projection of the position onto the vector
$\vN$, given by,
\begin{equation}
\vN=\vK+(\alpha/Q)\vJ,\qquad
\vN\cdot\vx=J^2-Q^2=\const,
\label{Nvector}
\end{equation}
implying that the trajectory lies in a plane perpendicular to $\vN$. The motion is,
therefore, a \emph{conic section}. Careful analysis would show that the trajectory is an ellipse,
a parabola, or a hyperbola, depending on the energy being smaller, equal or larger as $\beta$  \cite{Feher}. In particular, for
sufficiently low energies, the nuclear motion remains bounded.

The conserved vectors $\vJ$ and $\vK$ satisfy, furthermore,
the commutation relations
\begin{equation}\begin{array}{lll}
\big\{J_i,J_j\big\}&=&\epsilon_{ijk}J_k,
\\[6pt]
\big\{J_i,K_j\big\}&=&\epsilon_{ijk}K_k,
\\[6pt]
\big\{K_i,K_j\big\}&=&-2(H-\beta)\epsilon_{ijk}J_k,
\end{array}
\label{JKcommrel}
\end{equation}
with Casimir relations,
\begin{equation}
\vJ\cdot\vK=-\alpha\,Q,
\qquad
K^2=2(H-\beta)(J^2-Q^2)+\alpha^2. 
\label{casimirs}
\end{equation}
Normalizing $\vK$ 
 we get, therefore 
an $\SO(3)/ {\rm E(3)}/ \SO(3,1)$ dynamical symmetry, depending on the energy being smaller/equal/larger as $\beta$ \footnote{The dynamical symmetry actually extends
to an isospin-dependent representation of $\SO(4,2)$
\cite{HWY}.}.

We emphasize that the fine-tuned inverse-square term is necessary to overcome the obstruction in solving the constraint equation; without it, no Runge-Lenz vector would exist.

\section{Motion in self-dual monopole field}\label{SDSec}

The physical interpretation of the previous result is the following \cite{Feher}.  For large $r$, dropping exponentially decreasing terms,  
the  field of a {\it self-dual} non-Abelian monopole of  
charge $m$ is asymptotically  that of Wu-Yang, (\ref{WYmonopA}), augmented with a ``hedgehog'' Higgs field,
\begin{equation}
\Phi^a=\varphi\,\hat\Phi^a\ , 
\qquad \varphi=1-\frac{m}{r}\ ,\qquad\hat\Phi^a=\frac{x^a}{r}\ .
\end{equation}

The equations of motion of an isospin-carrying particle in the combined gauge- plus scalar field can conveniently be found by adding a fictitious fourth spatial dimension, $x^4$,
and putting $A_4^a=\Phi^a$ \cite{Feher}. Then $F_{i4}^a=D_j\Phi^a$, and the equations  (\ref{Wongeq}) yield
\begin{equation}\left\{
\begin{array}{lll}
\dot{\pi}_4&=&I^aD_j\Phi^a\dot{x}^j,
\\[8pt]
\dot{\pi}_i&=&-I^a\big(F^a_{ij}\dot{x}^j+D_i\Phi^a\dot{x}^4\big),
\\[8pt]
\dot{I}^a&=&\epsilon_{abc}I^b\big(A_j^c\dot{x}^j+\Phi^c\dot{x}^4\big).
\end{array}\right.
\label{4DWongeq}
\end{equation}
For the Wu-Yang gauge field, in particular, the direction field, $\hat\Phi^a$, is
covariantly constant, the non-Abelian field strength is parallel to $\hat\Phi^a$,
and the projection of the isospin onto $\hat\Phi^a$ is the conserved electric charge,
\begin{equation}
F^a_{ij}=f_{ij}\hat\Phi^a,
\qquad
D_j\hat\Phi^a=0,
\qquad
Q=I^a\hat\Phi^a.
\end{equation}
Here $f_{ij}$  
 is the (scalar) field strength of a Dirac monopole.
The upper equation in (\ref{4DWongeq}) is solved therefore by
\begin{equation}
\dot{x}^4=\pi_4=Q\varphi+Q_1
\end{equation}
where $Q_1$ is another constant of the motion.
Inserting $\dot{x}^4$ into the second equation in (\ref{4DWongeq}) yields,
at last, the generalized equations of motion in the self-dual field \cite{Feher,Duval:1982bb},
\begin{equation}
\dot{\pi}_i=-Qf^a_{ij}\dot{x}^j-\p_jV,\quad
V=\frac{1}{2}Q^2\varphi^2-QQ_1\varphi,
\label{modLoreq}
\end{equation}
(We observe that the third equations in (\ref{Wongeq}) and in (\ref{4DWongeq})
are also the same).
These equations derive, using the same Poisson bracket structure (\ref{PBracket}) as  above and a Hamiltonian (\ref{cHam}), with a potential of the form
(\ref{goodpot}).
Our results confirm, hence, the Kepler-type dynamical symmetry
in the asymptotic field of a self-dual monopole -- or equivalently, in an Abelian monopole field with 
self-dual scalar potential \cite{Feher}.

\section{Diatomic molecule}\label{DiatSec}

In Ref. \cite{MSW} Moody, Shapere and Wilczek 
have shown that, in the
Born-Oppenheimer approximation, nuclear motion in a diatomic molecule can be described by the effective  non-Abelian gauge field and Hamiltonian, 
\begin{eqnarray}
&&\begin{array}{l}
A_r=0,\qquad
A_\theta=\displaystyle\frac{1}{2}
\left[\begin{array}{cc}
0&-\kappa e^{i\phi}
\\[6pt]
-\kappa e^{-i\phi}&0
\end{array}\right],
\\[20pt]
A_\phi=\displaystyle\frac{1}{2}\left[\begin{array}{cc}
1-\cos\theta&-i\kappa e^{i\phi}\sin\theta
\\[6pt]
i\kappa e^{-i\phi}\sin\theta&-1+\cos\theta
\end{array}\right],
\end{array}
\label{DaVpot}
\\[6pt]
&&H=\frac{1}{2}\vpi^2+V,
\end{eqnarray}
where $\kappa$ is a real parameter.
The field strength resembles that of monopole aligned into the third internal direction,
$
F_{\theta\phi}=(1-\kappa^2)\sin\theta\,T_3,
$
except for the parameter $\kappa$ being unquantized.  The potential (\ref{DaVpot}) is
that a Wu-Yang [i.e., an imbedded Dirac] 
 monopole of unit charge when 
$\kappa=0$; for other values of $\kappa$, it is a truly non-Abelian configuration --- except for
$\kappa=\pm1$, when the field strength vanishes and (\ref{DaVpot}) is a gauge transform of the vacuum.

Our first step is present the field in a more convenient
form,
\begin{eqnarray}
\widetilde{A}_i^{\;a}=(1-\kappa)\epsilon_{iaj}\,
\frac{x_j}{r^2},
\quad 
\widetilde{F}^{\;a}_{ij}=(1-\kappa^2)\epsilon_{ijk}
\frac{x_kx_a}{r^4},
\label{diatfields}
\end{eqnarray}
which can be achieved by applying a suitable gauge transformation  
\cite{Jdiat}.
 
Turning to the conserved quantities, we note  that, 
when $\kappa\neq0$, the used-to-be electric charge,  $Q$ in (\ref{echarge}), is
\emph{not more covariantly conserved}  in general \footnote{
An exception is when the isospin is aligned into the radial direction, as seen from $(I^j-Q{x_j}/{r})x_j=0$.
A detailed calculation shows that the equation
$D_jQ=0$ can only be solved  when $\kappa=0,\pm1$.} , 
\begin{equation}
\big\{H,Q\big\}=-\vpi\cdot \vD Q, 
\quad
D_jQ=\frac{\kappa}{r}\left(I^j-Q\frac{x_j}{r}\right).
\label{noecharge}
\end{equation}
Nor is $Q^2$  conserved,
 $\big\{H,Q^2\big\}=-2\kappa Q(\vpi\cdot\vD Q)$. Note for further reference that,
unlike $Q^2$, the length of isospin, $I^2$, \emph{is} conserved, $\{H,I^2\}=0$.

The gauge field (\ref{DaVpot}) is rotationally symmetric
and an isospin-carrying particle moving in it
admits a conserved angular momentum \cite{MSW,Jdiat}. Its form is, however, somewhat unconventional, and we re-derive it, therefore, in detail.

Our starting point is the first-order condition in (\ref{constraints}). We take first $V=0$; then this is the only condition.
 Evaluating the r.h.s. with 
$F_{jk}^a$ as given in (\ref{diatfields}), the equation to be solved is
\begin{equation}
D_iC=(1-\kappa^2)\frac{Q}{r}\left((\vn\cdot\hx)\frac{x_i}{r}-n_i\right).
\label{diatC}
\end{equation}
In the Wu-Yang case, $\kappa=0$,  this equation was solved by
$C=-\vn\,\cdot\,Q{\hx}$. But for $\kappa\neq0$ the electric charge, $Q$, is not conserved, and using (\ref{noecharge}), as well as the relations
\begin{eqnarray}\begin{array}{ll}
D_iI^j&=
\displaystyle\frac{(1-\kappa)}{r}\left(Q\,\delta_{ij}-{I^i\frac{x_j}{r}}
\right)\ ,
\\[10pt]
D_i\left(Q\,\vn\cdot\displaystyle{\hx}\right)&=
\displaystyle\frac{Q}{r}\left(n_i+(\vec{n}\cdot\hat{r})
\big(\kappa I_i-(1+\kappa)\frac{r_i}{r}\big)\right)
\\[12pt]
I^aF_{ij}^a&=(1-\kappa^2)\,Q\,\displaystyle\frac{\epsilon_{ijk}x_k}{r^3}\ ,
\end{array}
\label{handyform}
\end{eqnarray}
 we find,
$$
(\kappa-1)D_i\left(Q\,\vn\cdot\hx\right)
=\kappa D_iI^jn_j+
(1-\kappa^2)\left((\vn\cdot\hx)\frac{x_i}{r}-n_i\right).
$$
Comparing with (\ref{diatC}) allows us to infer that,
\begin{equation}
C=-\vn\cdot\Big((1-\kappa)\,Q\,{\hx}+\kappa\vI\Big).
\label{JC}
\end{equation}
The conserved angular momentum is, therefore,
\begin{eqnarray}
\vJ&=&\vx\times\vpi-\vW,
\\[6pt]
\vW&=&(1-\kappa)\,Q\,\hx+\kappa\vI\ =
Q\,\hx+\kappa(\hx\times\vI),
\label{diatangmom}
\end{eqnarray}
consistently with the  results in \cite{MSW,Jdiat,Lee}.  
Comparison with (\ref{WYC}) yields the ``replacement rule'' 
\begin{equation}
Q\,\hx\to\vW.
\label{rule}
\end{equation}
For $\kappa=0$ we recover the Wu-Yang expression, (\ref{WYangmom}).
Eliminating $\vpi$ in favor of $\vp=\vpi+\vA$ allows us to rewrite the 
total angular momentum as
\begin{equation}
\vJ=\vx\times\vp-\vI\, ,
\end{equation}
making manifest the celebrated
``spin from isospin term'' \cite{Jackiw:1976xx}. 

Restoring the potential, we see that, again due to the non-conservation
of $Q$, 
 $D_jV\neq0$ in general. The zeroth-order condition $\vC\cdot\vD V=0$ in (\ref{constraints}) is, nevertheless, satisfied if $V$ is a radial function independent of $\vI$,
$V=V(r)$, since then $\vD V=\vec{\nabla} V$, which is perpendicular to infinitesimal rotations, $\vC$. 
 Alternatively, a direct calculation, using the same formulae
(\ref{noecharge})-(\ref{handyform}), allows us to
confirm that $\vJ$ 
commutes with the Hamiltonian, $\{J_i,H\}=0$.

Multiplying (\ref{diatangmom}) by $\hx$ yields, once again, the  relation  (\ref{cone}) i.e.  $\vJ\cdot\hx=-Q$, the same as in the Wu-Yang case.
This is, however, less useful as
before, since $Q$ is not a constant of the motion so that the angle between  $\vJ$ and the
radius vector, $\vx(t)$, is not more a constant.
The components of the angular momentum (\ref{diatangmom})
close, nevertheless, to $\SO(3)$,
$
\big\{J_i,J_j\big\}=\epsilon_{ijk}J_k.
$ 
 
Turning to quadratic conserved quantities, we have been searching for a Runge-Lenz vector for diatomic molecules
associated with the Killing tensor (\ref{RLKilling}).
Despite a promising start (summarized in the Appendix)
we failed to find such an additional conserved  quantity.

We should remark, however, that even if we succeeded to integrate
(\ref{diatDpot}), the resulting potential would \emph{break the rotational invariance}. The zeroth-order condition in (\ref{constraints}) requires
in fact the $D_jV$ be \emph{perpendicular} to $\vC$. But the
$\vC$ of angular momentum and the one appropriate for the Runge-Lenz vector,
namely the infinitesimal rotation in (\ref{rotKilling}) and (\ref{diatvC}), respectively, have different orientations, so that
the  two conditions can not be simultaneously satisfied.

\section{Relation to the Forg\'acs-Manton-Jackiw approach}\label{Fo-Ma-Ja}

Moody et al. \cite{MSW} found the correct
expression, (\ref{diatangmom}), for $\kappa=0$
but, as they say it, ``they are not aware of a canonical derivation
when $\kappa\neq0$''. 
Our construction here is an alternative to
that of  Jackiw \cite{Jdiat}, who obtained 
it using the method of Ref. \cite{JaMa}.
In his approach, based on the study of symmetric gauge fields \cite{FoMa},
each infinitesimal rotation, (\ref{rotKilling}),
is a symmetry of the monopole in the
sense that it changes the potential by a surface
term. Equivalently,
\begin{equation}
C_i F_{ij}=\cD_j\Psi_i 
\label{symmgcond}
\end{equation}
for some Lie algebra valued field $\Psi$. In this equation,
 $\cD$ is the gauge-covariant derivative $\cD_j\Psi=\p_j\Psi-[A_j,\Psi]$. Identifying the Lie algebra  with
 $\IR^3$ identifies $\Psi$ with a vector $\vPsi$,  the
 gauge-covariant derivative becoming the covariant derivative
 of van Holten in Eqn. (\ref{covder}). The rule is simply to replace the generator
 $T_a$ with the components of the isospin vector, $I^a$.
Under this transformation, the symmetry condition
(\ref{symmgcond}) becomes precisely the first-order condition  
in (\ref{constraints}) that a rotation has to satisfy.
Accordingly, in the diatomic case, eqn.  (\ref{symmgcond}) is solved by
\begin{equation}
\Psi=\left((1-\kappa)(\vn\cdot\hx)\frac{x_a}{r}+\kappa\, n_a\right)T_a\,,
\label{FMsym}
\end{equation}
consistently with the indicated correspondence 
$\Psi\leftrightarrow\vPsi$. It represents the response of the (symmetric) gauge field to a space-time
transformation and appears as the contribution coming from the gauge field to the associated conserved quantity
\cite{JaMa, DHInt}; it is needless to say that the 
same conserved quantities are obtained in both approaches.

Taken individually, each rotation generator is a symmetry. If we have several generators, an additional consistency condition is required, namely \cite{JaMa,FoMa}
\begin{equation}
\vX^i_{\vn}\vX^j_{\vm}F_{ij}=\Psi_{\vX_{\vn}\times\vX_{\vm}}
-\left[\Psi_{\vX_{\vn}},\Psi_{\vX_{\vm}}\right],
\label{GensymmGroupcond}
\end{equation}
where $\vX_{\vn}$ and $\vX_{\vm}$ are two infinitesimal rotations 
 around $\vn$ and $\vm$, respectively. Writing the l.h.s.
 as $r^2(\vn\,\vm\,\hx)(\vB\cdot\hx)$ where $\vB=(B_i)$ is the Lie algebra-valued vector 
$B_i^a=\epsilon_{ijk}F_{jk}^a$
 allows us to evaluate it, to find, in the diatomic case,  
\begin{equation}
(1-\kappa^2)\big(\hx\,\vn\,\vm\big)\displaystyle\frac{x_a}{r}\,T_a.
\label{lhs}
\end{equation}
On the other hand, we find
\begin{eqnarray}
\Psi_{\vn\times\vm}&=&\big(\hx\,\vn\,\vm\big)\displaystyle\frac{x_a}{r}T_a
\nonumber
\\[4pt]
&&+
\kappa\left[\hx\times(\vn\times\vm)\times\hx\right]_aT_a.
\label{Psinm}
\\[8pt]
-\left[\Psi_{\vX_{\vn}},\Psi_{\vX_{\vm}}\right]&=&
-\kappa^2\big(\hx\,\vn\,\vm\big)\displaystyle\frac{x_a}{r}T_a
\nonumber
\\[4pt]
&&
-\kappa\left[\hx\times(\vn\times\vm)\times\hx\right]_aT_a
\label{PsinPsim}
\end{eqnarray}
so that (\ref{GensymmGroupcond}) is indeed satisfied.

\section{Discussion}\label{DiscSec}

We derived the equations of motion in a Yang-Mills-Higgs background
 from those in a pure gauge field, by
viewing the Higgs field as the fourth component of the Yang-Mills field
on an extended space. 
It is worth mentioning that van Holten's algorithm could be extended 
to four dimensions. Then the electric charge could be seen  as associated with an internal symmetry generated by the covariantly constant direction field $\hat\Phi^a$ \cite{HR}; the second charge, 
$Q_1$ in \ref{SDSec}, corresponds in turn to the ``vertical'' Killing vector which points into the
fourth direction. Similarly, the Kepler potential
could be viewed as a component of the extended metric, and the Killing tensor
(\ref{RLKilling}) can be lifted to extended space, to yield the Runge-Lenz vector directly \cite{DGH}. 

Our results have a nice interpretation in terms of fiber bundles \cite{HR}.
The $\SU(2)$ gauge field is a connection form defined on a  bundle
over $3$-space. For Wu-Yang, this bundle reduces to an
$U(1)$ bundle, namely that of a Dirac monopole of unit charge.
The projection of the isospin onto a direction field, $\hat\Phi^a,\ |\hat\Phi|=1$, 
is a conserved
charge if and only if $\hat\Phi^a$ is \emph{covariantly constant}. But this amounts
precisely to saying that the $\su(2)$ connection living on the (trivial) 
$\SU(2)$ bundle reduces to the $U(1)$ Dirac monopole bundle. This explains why
the  electric charge (\ref{echarge}) is conserved in the Wu-Yang case: the latter is in fact an imbedded Abelian monopole. The non-Abelian
equations of motion (\ref{Wongeq}) reduce, accordingly, to those of a 
charged particle in an electromagnetic field.

We remark that although our investigations have been purely classical,
there would be no difficulty to extend them to a quantum particle.
In the self-dual Wu-Yang case, the $\SO(4)/\SO(3,1)$ dynamical
symmetry allows, in particular, to derive the bound-state spectrum 
 and 
the $S$-matrix group-theoretically, using the algebraic relations (\ref{JKcommrel})-(\ref{casimirs}) \cite{Feher}.

The effective field of a diatomic molecule provides us with an
interesting generalization. 
For $\kappa\neq0,\pm1$, it is truly non-Abelian, i.e., \emph{not reducible} to one on an $U(1)$ bundle. 
No covariantly constant direction field, and, therefore,
\emph{no conserved electric charge} does exist in this case.

The field is nevertheless radially symmetric, but the conserved angular momentum, (\ref{diatangmom}), has  a non-conventional form  \cite{Jdiat}.

In bundle terms, the action of a symmetry generator
can be lifted to the bundle so that it preserves
the connection form which represents the potential.
But the group structure may not be conserved;
this requires another, consistency condition, namely 
(\ref{GensymmGroupcond}),
 which may or may not be satisfied. In the diatomic case, it is not satisfied when  $\kappa\neq0,\,\pm1$. 
 
 Is it possible to
redefine the ``lift" so that the group structure
will be preserved~? In the Abelian case, the
answer can be given in cohomological terms \cite{DHInt}. If this obstruction 
does not vanish, it is only a {\it central
extension} that acts on the bundle.

In the truly non-Abelian case, the  consistency condition
involves the covariant, rather than ordinary derivative and covariantly constant sections 
only exist in exceptional cases -- namely when the
bundle is reducible. Thus, only some
(non-central extension) acts on the bundle.

It is worth noting that for $\kappa\neq0$ the configuration 
(\ref{DaVpot}) does not satisfy the vacuum Yang-Mills equations; it
only satisfies them with a suitable (conserved) current \cite{Jdiat},
\begin{eqnarray}
\cD_iF_{ik}=j_k,
\qquad
\vec{\jmath}=\frac{\kappa(1-\kappa^2)}{r^4}\,\vx\times\vec{T}\,,
\label{current}
\end{eqnarray}
Interestingly, this current can also be produced by a hedgehog Higgs field,
\begin{eqnarray}
 j_k=\big[\cD_k\Phi,\Phi\big],
\qquad
\Phi^a=\frac{\sqrt{1-\kappa^2}}{r}\,\frac{x_a}{r}\;.
\end{eqnarray}

We have not been able to derive
a Runge-Lenz vector for diatomic molecules, except for $\kappa=0$.£

Let us emphasize that the derivation of the 
non-Abelian field configuration (\ref{diatfields})
from molecular physics \cite{MSW} indicates  that  our
analysis may not be of purely academic
interest. The situation could well be analogous to what
happened before with the non-Abelian
Aharonov-Bohm experiment, first put forward
and studied theoretically \cite{NABAWY,NABAHPA},
but which became recently accessible experimentally, namely
by applying laser beams to cold atoms \cite{Ohberg}.
A similar technique can be used also to create monopole-type fields  \cite{Ohberg2}.

\begin{acknowledgments}
One of us (JPN) is indebted to the {\it R\'egion Centre} for a doctoral scholarship and to the {\it Laboratoire 
de Math\'ematiques et de Physique Th\'eorique} of Tours University  for hospitality extended to him.
We would like to thank Professor P. \"Ohberg for tutoring us on cold atoms.
Discussion with Professor P. Forg\'acs  and
correspondence with Prof. A. Shapere
are gratefully acknowledged. 
\end{acknowledgments}


\appendix{\section{Searching for a Runge-Lenz vector for
diatomic molecules}} 

We carried out a promising, but unsuccessful, search for finding 
a conserved Runge-Lenz vector in the effective gauge field of a
diatomic molecule. Inserting the Killing tensor (\ref{RLKilling}) 
 into the 2nd-order equations in
 (\ref{constraints}) yields, after a calculation similar to the one 
 sketched for the angular momentum,
\begin{eqnarray}
\vC = \vn\times\vW, 
\label{diatvC}
\end{eqnarray}
which is, once again, consistent with the ``replacement rule'' 
(\ref{rule}), cf. (\ref{WYvC}) 

Our next step is to try to identify a ``good'' potential.  For this,
we observe that
the first-order constraint in (\ref{constraints})  
can be written as,
\begin{eqnarray}
D_iC&=&(1-\kappa^2)\,\frac{Q^2}{r^2}\Big(
(1-\kappa)(\vn\cdot\hx)\frac{x_i}{r}-n_i\Big)
\label{dconst1}
\\[6pt]
&+&\kappa(1-\kappa^2)\displaystyle\frac{Q}{r^2}
(\vn\cdot\hx)I^i+C_{ij}D_jV.
\nonumber
\end{eqnarray}
In the WY case, $\kappa=0$, our clue has been to remove the first term 
by a fine-tuned inverse-square term in the potential,
$V=V_0+V_1$, $V_0=Q^2/2r^2$.
We note now that this is once again possible, namely choosing,
 \begin{equation}
\vD V_0=-(1-\kappa^2)\frac{Q}{r^3}\,
\vW, 
\label{diatDpot}
\end{equation}
 once again consistently with our rule (\ref{rule}). 
Assuming that such a potential does exist, 
$C_{ij}D_jV_0$
cancels the upper term in (\ref{dconst1}) [but contributes others],
leaving us with
\begin{eqnarray}
D_iC
=\kappa(1-\kappa^2)\displaystyle\frac{Q}{r^2}\big(\vn\times(\hx\times\vI)
\big)_i+C_{ij}D_jV_1.
\label{dconst1bis}
\end{eqnarray}

Our remaining task would now be to integrate the equations (\ref{diatDpot})
and (\ref{dconst1bis}) --- that we have not been able to do yet
in general.
We could do it 
 for $\kappa=0$, though,  allowing us to recover the
 results previously found in Section \ref{WYSec}. The electric charge, $Q$, is now conserved, and eqn. (\ref{diatDpot}) can now be integrated,
\begin{equation}
\vD V_0=-\frac{Q^2}{r^3}\,\hx
\quad
\Rightarrow\quad V_0=\frac{Q^2}{2r^2}.
\label{k0diatDpot}
\end{equation}
Now assuming that $V_1$ only depends on $r$, $V=V_1(r)$,
 $\vD_iV_1=\vnabla V_1$ is radial. On the other hand,
using (\ref{radCij})
 Eqn. (\ref{dconst1bis}) reduces to
\begin{equation}
D_iC=C_{ij}\p_jV_1=-r^2V_1'\vnabla(\vn\cdot\hx)
\end{equation}
where $V_1'=dV_1/dr$, which can be solved 
by 
\begin{equation}
V_1=\frac{\alpha}{r}+\beta,
\qquad
C=\alpha(\vn\cdot\hx),
\end{equation}
consistently with what we found before in (\ref{goodpot}).
Some more calculations show, furthermore, that this is the only
possibility.

In the MacIntosh-Zwanziger (and hence the self-dual Wu-Yang) case, the arising of the ``fine-tuned'' potential
can also be understood as follows. Decomposing
the momentum into radial and transverse components,
$
(\vpi)^2=(\vpi\cdot{\hx})^2+(\vpi\times{\hx})^2
=\pi_r^2+{L^2}/{r^2}  
$ 
and using
$\vL^2=\vJ^2-Q^2$
 allows us to present the Hamiltonian as 

$$
H=H_0+V=\frac{1}{2}p_r^2+\frac{\vJ^2}{2r^2}
-\frac{Q^2}{2r^2}+V.
$$
Therefore, choosing the potential as in (\ref{goodpot})
cancels the centrifugal term $Q^2/2r^2$, leaving us with,
\begin{equation}
H=\frac{1}{2}p_r^2+\frac{\vJ^2}{2r^2}
+\frac{\alpha}{r}+\beta,
\label{effKep}
\end{equation}
which describes an effective Kepler-type problem.
For diatoms,  the Hamiltonian is, instead 
\begin{eqnarray}
&H=\displaystyle\frac{1}{2}p_r^2+ \displaystyle\frac{J^2}{2r^2}+
\nonumber
\\[6pt]
&\left\{\displaystyle\frac{-(1-\kappa)^2Q^2+\kappa^2I^2+2\kappa\vJ\cdot\vI}{2r^2}\right\}
+V.
\label{diatraddec}
\end{eqnarray}
One would be tempted to chose the potential so that it cancels the
curly bracket leaving us, once again, with a Kepler-type
radial Hamiltonian. This does not work, however, since the
covariant derivative of such a $V$ is not perpendicular to the vector $\vC$
in (\ref{diatvC}).

Eliminating the total angular momentum, $\vJ$, in favor of the orbital one, $\vL=\vx\times\vpi$, and
putting  $Q^2=I^2=1/4$ would yield  the decomposition \# (21) of  Jackiw \cite{Jdiat}. This is, however, only legitimate when
 $\kappa=0$, since  $Q^2$ is not conserved for $\kappa\neq0$.

We mention that we can also solve our equations (\ref{diatDpot}) and
(\ref{dconst1bis}) in another particular case, namely when
the \emph{isospin is radially aligned}, 
$\vI=I_0\hx.$ Then $Q=I_0$ is conserved.
 This condition
is very restrictive, however: the equations of motion (\ref{Wongeq}) imply that if isospin alignment is required as an
initial condition at $t=0$, it only remains satisfied for 
all $t$ if the space-time motion is radial.

When the isospin is aligned,  the 
curly bracket in (\ref{diatraddec})  becomes simply $-Q^2/2r^2$, and we are back in the Wu-Yang case --- but only for radial motions  such that the isospin is also radial,  $\vI=I_0\hx$.

For $\kappa=0$, the integrability of the equations (\ref{dconst1bis})
can also be studied as follows. By (\ref{covdercomm}), we must have
\begin{equation}
-\epsilon_{abc}I^aF_{ij}^b\frac{\partial C}{\partial I^c}
=(n_ix_j-n_jx_i)\bigtriangleup V_1.
\label{WYintcond}
\end{equation}
Assuming that $C$ only depends on $\vx$, the l.h.s. vanishes, and this condition
merely requires
\begin{equation}
\bigtriangleup V_1=0
\qquad\Rightarrow\qquad
V_1=\frac{\alpha}{r}+\beta,
\end{equation}
as we found it before.

\end{document}